\documentclass[amsmath,title,amssymb,aps,prl,superscriptaddress,twocolumn,floatfix,title]{revtex4-2}
\usepackage{float}
\usepackage{xcolor}
\usepackage{color}
\usepackage{graphicx}
\usepackage{dcolumn}
\usepackage{bm}
\usepackage{color}
\usepackage{amsmath}
\usepackage{amsfonts}
\usepackage[colorlinks]{hyperref}
\usepackage{graphicx}
\usepackage{dcolumn}
\usepackage{bm}
\usepackage{cleveref}

\hypersetup{
citecolor={blue},
urlcolor={blue}
}

\usepackage{cleveref}
\crefname{appendix}{Appendix}{Appendices}
\crefname{equation}{Eq.}{Eqs.}
\crefname{figure}{Fig.}{Figs.}
\crefname{table}{Table}{Tables}
\crefname{section}{Section}{Sections}
\crefname{mythe}{Theorem}{Theorems}
\crefname{mydef}{Definition}{Definitions}
\newcommand{\dummylabel}[2]{\def\@currentlabel{#2}\label{#1}}

\begin{document}
\title{Band-spin-valley coupled exciton physics in antiferromagnetic MnPS$_3$}

\author{Dan Wang}
\affiliation{State Key Laboratory of Low Dimensional Quantum Physics and Department of Physics, Tsinghua University, Beijing 100084, China}

\author{Haowei Chen}
\affiliation{State Key Laboratory of Low Dimensional Quantum Physics and Department of Physics, Tsinghua University, Beijing 100084, China}

\author{Yu Pang}
\affiliation{Shenzhen Geim Graphene Center $\&$ Shenzhen Key Laboratory of Advanced Layered Materials for Value-added Applications, Institute of Materials Research, Tsinghua Shenzhen International Graduate School, Tsinghua University, Shenzhen 518055, China}

\author{Xiaolong Zou}
\email{xlzou@sz.tsinghua.edu.cn}
\affiliation{Shenzhen Geim Graphene Center $\&$ Shenzhen Key Laboratory of Advanced Layered Materials for Value-added Applications, Institute of Materials Research, Tsinghua Shenzhen International Graduate School, Tsinghua University, Shenzhen 518055, China}

\author{Wenhui Duan}
\affiliation{State Key Laboratory of Low Dimensional Quantum Physics and Department of Physics, Tsinghua University, Beijing 100084, China}
\affiliation{Institute for Advanced Study, Tsinghua University, Beijing 100084, People's Republic of China}
\affiliation{Frontier Science Center for Quantum Information, Beijing 100084, People's Republic of China}

\begin{abstract}
The introduction of intrinsic magnetic order in two-dimensional (2D) semiconductors offers great opportunities for investigating correlated excitonic phenomena. Here, we employ full-spinor GW plus Bethe-Salpeter equation methodology to reveal rich exciton physics in a prototypical 2D N\'{e}el-type antiferromagnetic semiconductor MnPS$_3$, enabled by the interplay among inverted dispersion of the second valence band, spin-valley coupling and magnetic order. The negative hole mass increases the reduced mass of the lowest-energy bright exciton, leading to exchange splitting enhancement of the bright exciton relative to band-edge dark exciton. Notably, such splitting couples with spontaneous valley polarization to generate distinct excitonic fine structure between $K$ and $-K$ valleys, which dictate distinct relaxation behaviors. Crucially, magnetic order transition from N\'{e}el antiferromagnetic to ferromagnetic state induces significant quasiparticle band structure reconstruction and excitonic transitions modification, with low-energy optical excitations being exclusively contributed by majority-spin channel. These findings establish 2D antiferromagnetic semiconductors as an intriguing platform to study band-spin-valley coupled exciton physics.
\end{abstract}

\maketitle

\section{\label{sec:level1}I. INTRODUCTION}
Two-dimensional (2D) transition-metal dichalcogenides (TMDCs) exhibit strong light-matter interaction, leading to a wealth of fascinating excitonic physics, in particular, the distinctive valley-dependent excitonic responses. The degeneracy of conduction and valence band edges at inequivalent $K$ and $-K$ valleys enables the formation of exciton states with coupled spin and valley indices \cite{RN1,RN2,RN3,RN4,RN5,RE_review}. Moreover, large exchange interaction results in a rich landscape of bright and dark excitons \cite{RN6,RN7,RN8,RN9}, which determines their light-coupling efficiency. To harness the valley degree of freedom in practical applications, achieving valley polarization constitutes a critical prerequisite, as it permits selective carrier excitation from distinct valleys \cite{RN10,RN11,RN12}. This fundamental requirement underscores the necessity of expanding material exploration beyond conventional TMDCs, where novel valley-dependent excitonic properties can be accessed and engineered for advanced quantum technologies.

The discovery of 2D magnets CrI$_3$ \cite{RN13} and Cr$_2$Ge$_2$Te$_6$ \cite{RN14} has sparked interest in the excitonic properties of magnetic semiconductors. Giant excitonic and magneto-optical effects have been predicted in monolayer CrI$_3$ \cite{RN15} and CrBr$_3$ \cite{RN16}, demonstrating the critical role of spin degree of freedom in determining optical behaviors. Recently, a family of 2D intralayer antiferromagnetic transition metal phosphorus trichalcogenides (\textit{M}P\textit{X}$_3$, \textit{M} = Mn, Ni, Fe; \textit{X} = S, Se) has attracted significant attention, especially for their intriguing excitonic effects \cite{RN17,RN18,RN19}. Of particular interest is MnPS$_3$, a wide-bandgap compound that exhibits spontaneous valley polarization arising from its unique spin-valley coupled physics \cite{RN20}. Previous theoretical results reveal large exciton binding energies for MnPS$_3$ \cite{RN21,RN22}, exceeding those of TMDCs. Despite recent progress in understanding the magnetic properties and ground-state exciton binding energy of MnPS$_3$, the band-spin-valley coupled excitons and their correlations with magnetic order remain to be elucidated, which could offer a new platform for investigating exciton physics with strong correlation.

In this paper, we employ density functional theory with many-body perturbation [i.e., GW and Bethe-Salpeter equation (BSE)] to investigate the exciton fine structure of monolayer N\'{e}el-type antiferromagnetic MnPS$_3$, aiming to address how the excitons couple with spin and valley degrees of freedom. We identify an unusual bright exciton (B$_1$) consisting of an electron from the first conduction band and a hole from the second valence band with negative effective mass, leading to an increased excitonic reduced mass, and enhanced exchange splitting. The resulting excitonic exchange splitting reaches 695 meV, a magnitude ten times larger than those in TMDCs. Furthermore, a variety of optically forbidden states reside below B$_1$, exerting significant influences on the dynamics of optically accessible states. Notably, spin-valley coupling synergizing with the huge exchange splitting generates distinct excitonic fine structures between $K$ and $-K$ valleys, thereby driving valley-contrasting microscopic relaxation channels of optically excited excitons. Finally, upon switching the magnetic order to the ferromagnetic phase, we observe significant quasiparticle band reconstruction accompanied by a dramatic reduction in exciton binding energies. In the ferromagnetic phase, spin selection rules strictly forbid optical transitions between conduction band minimum (CBM) and valence band maximum (VBM), restricting bright exciton formation exclusively in the majority-spin channel.

\section{II. COMPUTATIONAL METHODS}
The atomic and electronic structures are obtained by using density-functional theory (DFT) as implemented in the VASP package \cite{RN23}. We use the generalized gradient approximation by Perdew-Burke-Ernzerhof (PBE) \cite{RN24,RN25} with an onsite Hubbard potential ($U=5$ eV) as an approximation of the exchange-correlation term. Heyd-Scuseria-Ernzerhof (HSE06) hybrid functional \cite{RN26} is employed to further validate our results. The ground-state wave functions and eigenvalues are calculated within a $k$-grid of $15\times15\times1$ and a plane-wave cutoff of 500 eV. The structures are relaxed until the total forces are less than 0.01 eV/\AA, and the convergence criterion for total energies is set to 10$^{-7}$ eV. Spin-orbit coupling (SOC) is fully considered in our calculations. 

The quasiparticle band structures and excitonic properties are calculated with the YAMBO code \cite{RN27,RN28}. Using the QUANTUM ESPRESSO package \cite{RN29}, the PBE approximation has served as a mean-field starting point. Single-shot G$_0$W$_0$ \cite{RN30} is carried out to access the quasiparticle energy with $21\times21\times1$ $k$-point grid to sample the Brillouin zone. A slab model is used with a vacuum layer larger than 15 \AA\ along the out-of-plane direction, and a truncated Coulomb interaction is adopted. A total of 1000 bands were used to ensure the band gap converges within 0.02 eV. Excitonic properties are obtained by solving the BSE \cite{RN31,RN32} on top of GW results with ten valence and ten conduction bands included to converge the optical spectra. Our convergence test is supplied in Table SI in the Supplemental Material \cite{SM} (and Ref. \cite{RN284} therein).

\section{III. RESULTS AND DISCUSSIONS}
\begin{figure*}
\centering
\includegraphics[width=14cm]{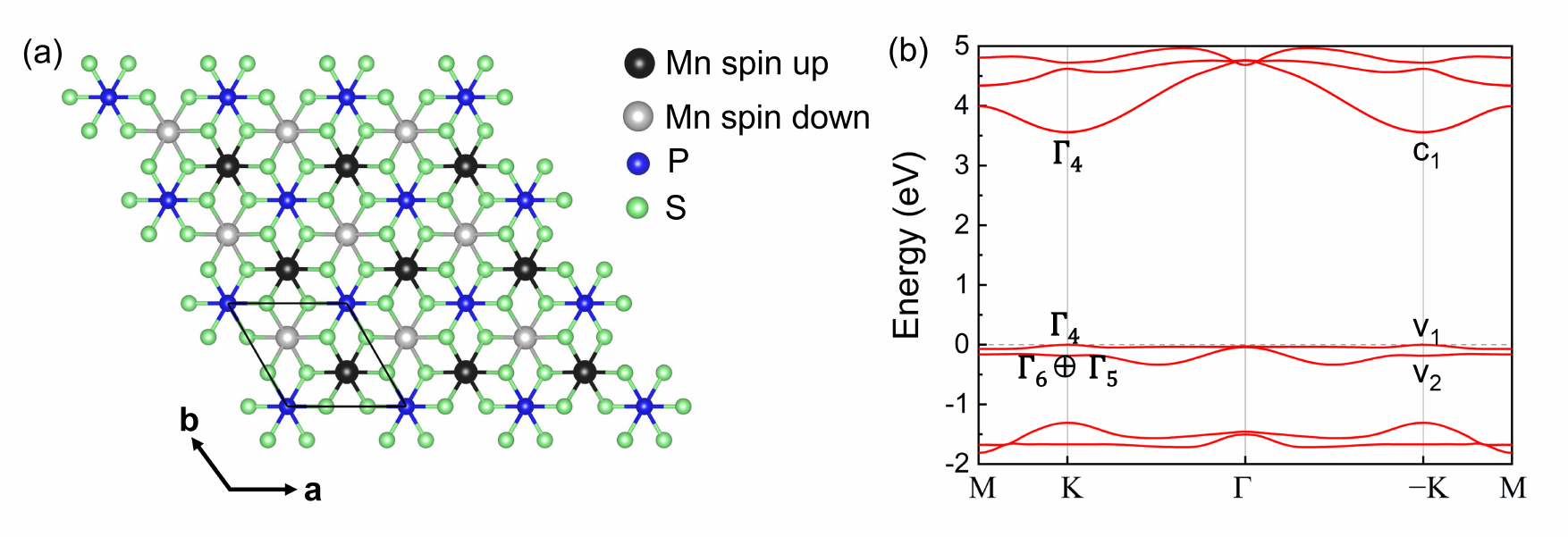}
\caption{(a) Crystal structure of monolayer MnPS$_3$. Spin arrangement of the Mn atoms in the N\'{e}el type AFM state is distinguished by black and gray spheres. The unit cell is outlined by solid black lines. (b) GW band structure of monolayer MnPS$_3$. The bands are labeled by corresponding irreducible spinor representations, and the energy of the valence band maximum is set to zero.}
\label{fig1}
\end{figure*}

MnPS$_3$ is an intralayer antiferromagnet with a N\'{e}el order of Mn$^{2+}$ magnetic moments, and the N\'{e}el temperature of the bulk material is $T_N$ = 78 K \cite{RN33}. Recent experiments reported that the antiferromagnetic (AFM) order persists down to monolayer \cite{RN34}. The atomic structure of monolayer MnPS$_3$ is presented in Fig. \ref{fig1}(a). Mn atoms are arranged in a honeycomb structure and sandwiched between two PS$_3$ tetrahedrons. The lattice structure belongs to $D_{3d}$ point group, and its N\'{e}el order phase corresponds to a magnetic point group $-3'm$ preserving the spatial-time reversal symmetry $(\mathcal{PT})$. Based on our optimized cells (calculated lattice constant a = 6.050 \AA\ versus experimental result of 6.077 \AA \cite{RN35}, the band structure calculated by the GW method is shown in Fig. \ref{fig1}(b). Although the Hubbard $U$ term affects the position of the valence band extrema and band edge curvatures (see Fig. S1 in the Supplemental Material \cite{SM}), we selected PBE as the starting point for our GW calculations. This choice is reasonable because GW and DFT + $U$ theories are both Hartree–Fock-like theories for localized states such as \emph{d} states where on-site Coulomb correlation is important \cite{RN37}. As seen from Fig. S1(d) \cite{SM}, the top of the valence band at the PBE level locates around the midpoint of the $K$-$M$ line, while it shifts to $K$ point at the GW level, which is consistent with PBE + $U$ results. We have also calculated band structure with HSE06 functional in Fig. S1(c) \cite{SM}, which shows similar dispersions with those obtained with GW and PBE + $U$ ($U=5$ eV). 

For monolayer MnPS$_3$, there exists spontaneous valley polarization between $K$ and $-K$ in the presence of SOC (see Tables SII and SIII \cite{SM}), which is coupled to the out-of-plane N\'{e}el AFM order \cite{RN20}. On the contrary, in the absence of the SOC, the band extrema at $K$ and $-K$ are degenerate (Fig. S1(e) \cite{SM}). Although SOC renormalizes the valley polarization, it does not alter the bandgap and band dispersion near the band edges significantly. The direct bandgap is 2.296 eV and 2.303 eV under PBE + $U$, whereas the GW quasiparticle bandgap including self-energy correction is 3.523 eV and 3.542 eV at the $K$ and $-K$ points, respectively. The energy gap differences between $\pm K$ are 6.4 and 19.0 meV at PBE + $U$ and GW levels, respectively. Moreover, to compare with available experimental data, the calculated GW gap (3.52 eV) is larger than the band gap of bulk MnPS$_3$ (3.0 eV \cite{RE_gap1979,RE_gap2021}), which is due to the reduced dielectric screening in low dimension.

\subsection{\label{sec:level2}A. Band-valley coupled excitons}
\begin{figure*}
\centering
\includegraphics[width=13cm]{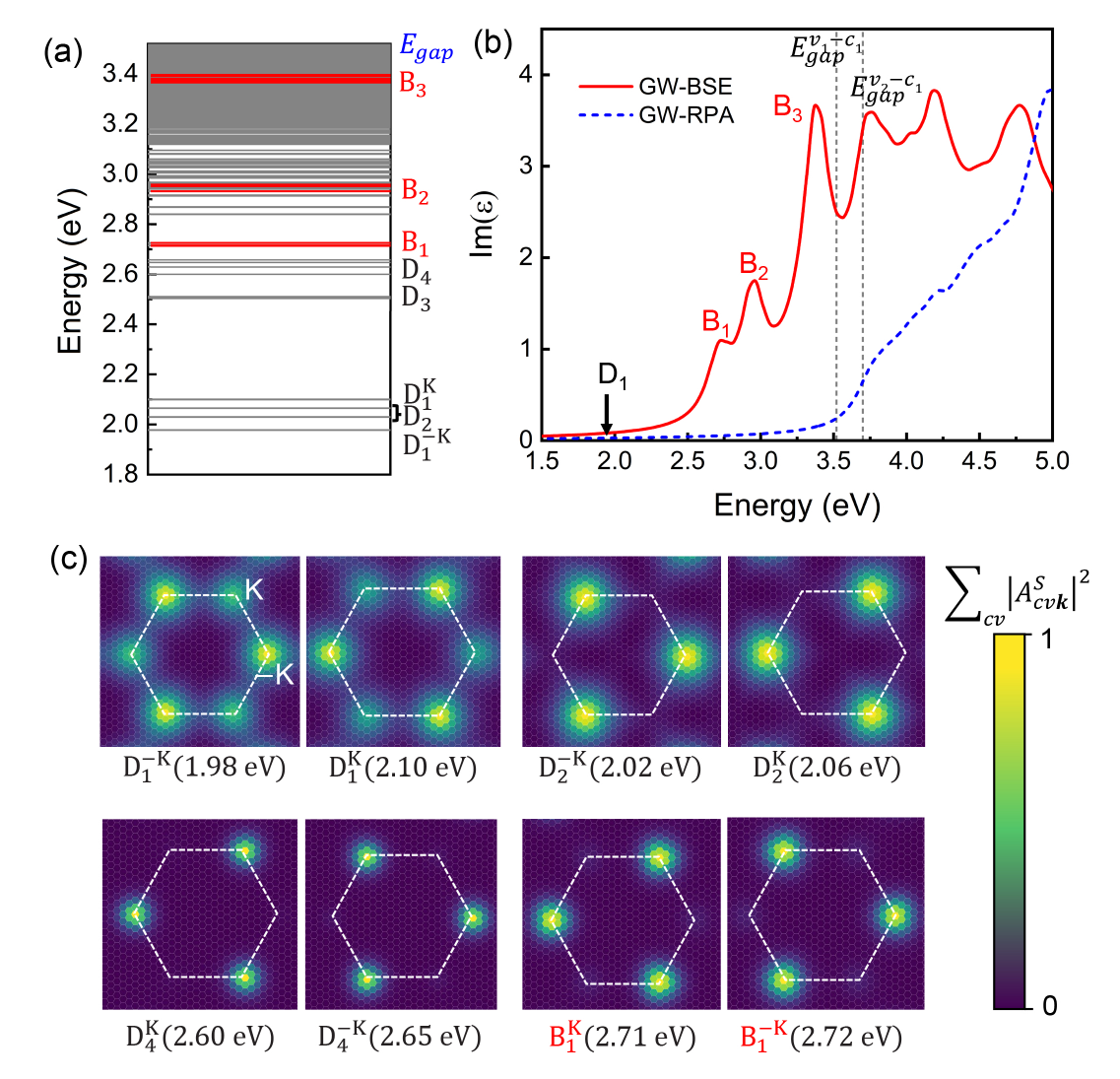}
\caption{(a) Exciton energy levels of monolayer MnPS$_3$ calculated using GW-BSE method. Optically bright exciton states are in red, while dark ones are shown in gray lines. The quasiparticle band gap is 3.52 eV. We label the bound exciton states with D for the lowest-lying dark states and B for the higher-lying bright states. (b) Absorption spectrum for in-plane polarized light with electron-hole interaction (GW-BSE, solid red line) and without electron-hole interaction (GW-RPA, dashed blue line).  The vertical black dashed line indicates the direct gap of v$_1$ to c$_1$ and v$_2$ to c$_1$ at $K$ point. (c) Exciton envelope functions in $k$-space of D$_1$, D$_2$, D$_4$ and B$_1$ excitons with their excitation energies \emph{E$_S$} labeled in brackets. The white dotted-line hexagon denotes the first Brillouin zone. v$_1$ to c$_1$ transition contributes to D$_1$ and D$_4$ excitons, while v$_2$ to c$_1$ transition contributes to D$_2$ and B$_1$.}
\label{fig2}
\end{figure*}

To investigate excitonic properties, we solve BSE with electron-hole interaction included \cite{RN31,RN32}: 
\begin{eqnarray}
A_{cv\bm{k}}^S(E_{c\bm{k}}-E_{v\bm{k}})+\displaystyle\sum_{c'v' \bm{k'}}A_{c'v'\bm{k'}}^S\langle cv\bm{k} |\hat{K}| c'v'\bm{k'} \rangle \nonumber \\ 
= A_{cv\bm{k}}^SE_S,
\end{eqnarray}
where $E_{c\bm{k}}$ and $E_{v\bm{k}}$ are quasiparticle energies of the conduction and valence bands, respectively. The eigenstate for an exciton state $S$ with excitation energy $E_S$ is expressed as $|{S} \rangle=\sum_{cv\bm{k}}A_{cv\bm{k}}^S|cv\bm{k} \rangle$, with $A_{cv\bm{k}}^S$ $k$-space envelope of the excitonic wave function. $\hat{K}=\hat{K}_d+\hat{K}_x$ is the electron-hole interaction kernel, consisting of an attractive direct screened term $\hat{K}_d$ and an exchange term $\hat{K}_x$. After solving BSE with spinor wave functions, a series of strongly bound dark and several bright exciton states with excitation energies $E_S$ below the quasiparticle bandgap ($E_{gap}$) are observed, as schematically shown in Fig. \ref{fig2}(a). Figure \ref{fig2}(b) shows the calculated in-plane polarized absorption spectrum of monolayer MnPS$_3$ without (dashed blue line labeled as GW-RPA) and with electron-hole interaction (solid red line denoted as GW-BSE). At GW-RPA level, the absorbance edge starts at around 3.7 eV, corresponding to the quasiparticle direct band gap between v$_2$ to c$_1$ at $\pm K$ valleys, while the optical transition from v$_1$ to c$_1$ is negligibly small. From group theory analysis, the symmetry of the states at $\pm K$ is characterized by the $-3’m$ magnetic point group and its unitary symmetry operations obeys the $C_{3v}$ point group, whose character table is shown in Table SV. The irreducible representations of low-energy bands are specified in Fig. \ref{fig1}(b), which is determined with IRVSP \cite{RN38}. The spinor representations of v$_1$ and c$_1$ transform according to $\Gamma_4$, while that of v$_2$ follows $\Gamma_5\oplus \Gamma_6$. The corresponding excitonic states with 1$s$ envelope function can be expressed as the direct product of $\Gamma_4\otimes\Gamma_4=\Gamma_1\oplus \Gamma_2\oplus \Gamma_3$ and $\Gamma_4\otimes(\Gamma_5\oplus \Gamma_6)=\Gamma_3$. It is noted the transition is dipole allowed between v$_1$ to c$_1$, for both out-of-plane and in-plane polarized light belonging to the $\Gamma_2$ and $\Gamma_3$ representations, respectively. However, the v$_1$ band around $\pm K$ point is mostly composed of the S atoms $p_x\pm ip_y$ states, and c$_1$ band originates from $d_{x^2-y^2}\pm id_{xy}$ states of Mn, according to the projected DOS plotted in Fig. S2 \cite{SM}. Accordingly, the overlap between electron and hole wave functions, which are all restricted in the single atomic Mn or S layer, is very small. This results in negligible interband transition dipole moment $\langle v_1\bm{k} |\bm{\hat{p}}| c_1\bm{k} \rangle$ for the in-plane polarized light. In contrast, the interband transition can slightly couple with $z$-polarized light, and gains some dipole oscillator strength. This is in consistent with the $z$-polarized absorption spectrum with an additional peak (denoted as D$_4$) below B$_1$ (originating from v$_2$ to c$_1$ transition as discussed below) shown in Fig. S3 \cite{SM}, but with one orders of magnitude weaker intensity than the first bright peak B$_1$. We also calculate the module square of transition matrix elements along high-symmetry lines shown in Fig. S4 \cite{SM}. Evidently, the interband transition from v$_1$ to c$_1$ shows tiny out-of-plane component around $\pm K$ valleys, while v$_2$ to c$_1$ transition exhibits strong in-plane component.
\begin{table}[htbp]
  \centering
  \caption{Calculated effective mass of the electron and hole at their respective bands as well as excitonic reduced mass for the exchange-split pairs D$_1$ and D$_4$ (denoted as $\mu_D$) and for the pairs D$_2$ and B$_1$ (labeled as $\mu_B$) at $\pm K$ valleys.}
  \begin{ruledtabular}
  \begin{tabular}{*{3}{c}}
    Mass($m_e$)  & $K$   & $-K$    \\ \hline
    v$_2$  & \textminus0.687  & \textminus0.645  \\
    v$_1$  & 1.084   & 1.087   \\
    c$_1$  & 0.555  & 0.567  \\
    $\mu_D$  & 0.361  & 0.366  \\
    $\mu_B$  & 2.888  & 4.689  \\
  \end{tabular}
  \end{ruledtabular}
\end{table}

Valley physics in MnPS$_3$ shows critical influences on its excitonic properties. In Fig. \ref{fig2}(c), we plot the normalized module square of exciton envelope functions in $k$-space. The ground-state dark excitons D$_1$ primarily localize around $-K$ valley, with minor contributions mixed from the $K$ valley. The lowest-energy dark excitons at $K$ valley D$_1(K)$ lies 0.12 eV higher than D$_1(-K)$. The different contributions of $-K$ and $K$ valleys to $k$-space envelope functions originate from the spontaneous valley splitting in MnPS$_3$, in stark contrast to excitons in monolayer TMDCs which are equally contributed by two valleys \cite{RN4,RN5}. 

Different from other 2D materials, the lowest bright exciton B$_1$ in MnPS$_3$ shows unique behaviors, due to its characteristic band dispersion. B$_1$ excition originates from holes in the second valence band (v$_2$) and electrons in the first conduction band (c$_1$) with conservation of the spin orientation, having an excitation energy of 2.71 eV and binding energy of 0.99 eV for $K$ valley. Our calculated B$_1$ exciton energy shows good agreement with the experimentally observed band-edge transition at 2.846 eV for multilayer MnPS$_3$ at 20 K \cite{RE_exp}. In contrast to v$_1$ with in-plane $p_x\pm ip_y$ character, v$_2$ is mostly composed of S $p_z$ orbital (Fig. S2 \cite{SM}) extending along out-of-plane direction. Since S and Mn atoms lie in different sublayers, the orbital overlap between v$_2$ and c$_1$ is expected to be strengthened, giving rise to the bright nature of B$_1$. Importantly, v$_2$ band has an opposite curvature compared to the v$_1$ at $\pm K$ valleys, corresponding to a negative effective mass hole. The picture of a stable exciton involving a negative-mass constituent electron or hole is rare. Most recently, this kind of exciton species have been observed in monolayer TMDCs \cite{RN39,RN40}, and identified as bound high-lying excitons (HXs) comprised of electrons with negative mass, which originates from the upper conduction bands. However, HXs in TMDCs are dark $p$-like and involves transitions from various nearby bands and $k$-points, which render this peak not detectable in conventional absorption or reflectance measurements. Differently, in monolayer MnPS$_3$, B$_1$ is the first bright excitons well-separated from other high-lying excitons, and hence could be observed by both optical absorption and luminescence. In addition, as depicted in Fig. \ref{fig3} (middle panel), electron-phonon scattering in a band with negative curvature should move holes higher in energy, away from the $\pm K$ valleys. This would generate a characteristic line shape composed of multiple equally spaced absorption peaks, attributed to the bright exciton and its phonon replicas. 

Under a simple effective-mass hydrogen model, such a bound electron-hole complex will be stable, as long as the reduced mass of the exciton $(\mu=(m_e^* m_h^*)⁄(m_e^*+m_h^*))$ remains positive \cite{RN39}. The hole mass $m_h^*$ of v$_2$ is negative and larger in magnitude than the positive electron mass $m_e^*$ of c$_1$, accordingly, the reduced mass of the B$_1$ is positive, as summarized in Table I. Meanwhile, the nearly parallel dispersion of v$_2$ and c$_1$ renders the possibility that excited electrons and holes travel together with a similar velocity in the same direction. We plot the real-space exciton wave functions of the low-energy excitons, with the hole fixed at the center of the six S atoms in Fig. S3(b) in Supplemental Material \cite{SM}. While D$_1$ and D$_4$ excitons (with holes from v$_1$) show Wannier behaviors extending over several unit cells, D$_2$ and B$_1$ excitons (with holes from v$_2$) resemble charge-transfer characteristics with the distribution of the electron density on a separated ring around the hole \cite{RN15,RN16,RN41}. These results establish the first observation of unconventional bright excitons involving negative-mass holes in an AFM vdW material, enriching exciton landscape in 2D materials.

\subsection{\label{sec:level2}B. Spin-valley coupled excitons}
\begin{figure*}
\centering
\includegraphics[width=16cm]{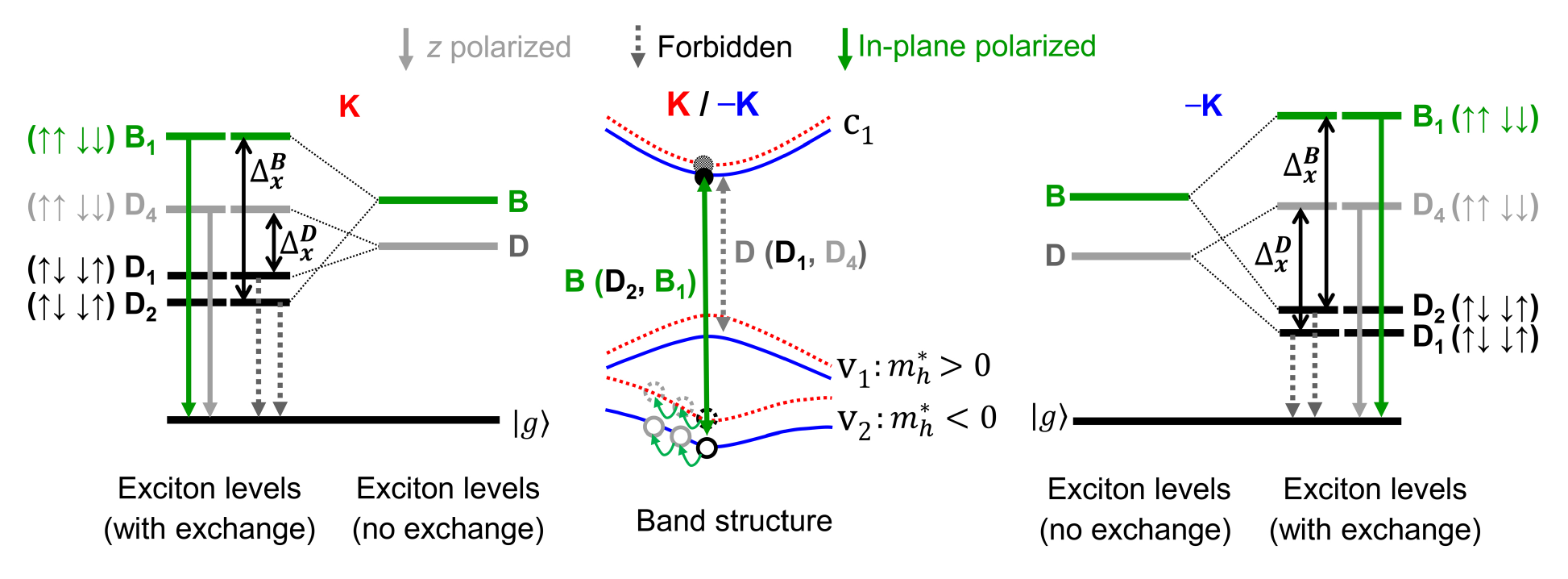}
\caption{Schematic illustration of band structure at $K$ valley in red dashed lines and $-K$ valley in blue solid lines (middle). Electrons from the conduction band minimum (c$_1$) recombine with holes from the second valence band (v$_2$) to emit light (green arrow), while the transition from c$_1$ to v$_1$ is largely forbidden. Fine structure splitting of low-energy excitons at $K$ and $-K$ is shown in the left and right panels, respectively. The number of short horizontal lines indicates the degeneracy of excitonic levels, and thus the degeneracy of all exciton states is two. $|g\rangle$ is the ground state and the arrows in parentheses represent electron-hole spin orientation. While D$_1$ and D$_2$ are dark spin-flip states, the spin-conserving D$_4$ and B$_1$ states are gray and bright states with out-of-plane and in-plane symmetries, respectively. Exchange interaction induced level splittings are labeled as $\Delta_x^D$ and $\Delta_x^B$.}
\label{fig3}
\end{figure*}

\begin{table*}[htbp]
  \centering
  \caption{Collection of excitonic exchange splitting energies for various 2D and 3D systems, together with our calculated exchange splitting energies $\Delta_x^D$ and $\Delta_x^B$ (as labeled in Fig. \ref{fig3}) in monolayer MnPS$_3$ at $K$ ($-K$) valley.}
  \begin{ruledtabular}
  \begin{tabular}{*{6}{c}}
    Materials  & $\Delta_x$(meV)   & Materials  & $\Delta_x$(meV) & Materials  & $\Delta_x$(meV)   \\ \hline
    CdSe       &	0.13 \cite{RN42} & (C$_4$H$_9$NH$_3$)$_2$PbBr$_4$ &	32 \cite{RN43} & MoS$_2$	& 14 \cite{RN8} \\
    InP         &   0.04 \cite{RN44} & (PEA)$_2$PbBr$_4$ &	27 \cite{RN45} & MoSe$_2$	& $-$1.4 \cite{RN8} \\
    GaAs/Al$_{1-x}$Ga$_{x}$As &	0.1 \cite{RN46} &  WS$_2$  &	55 \cite{RN8} & MnPS$_3$ $\Delta_x^D$ &	501 (671) \\
    GaAs/AlAs  &	0.15 \cite{RN47} & WSe$_2$	& 40 \cite{RN8} &  MnPS$_3$ $\Delta_x^B$	& 650 (695)  \\
  \end{tabular}
  \end{ruledtabular}
\end{table*}

Generally, for bound electron-hole pairs, the electron spin can be either parallel or antiparallel to the spin of the hole. It has been well-known that the spin degeneracy of the excitons in semiconductors is partly lifted by the electron-hole exchange interaction, which produces a fine structure of singlet and triplet excitons. Considering the weak spin-orbit coupling in MnPS$_3$, the singlet-triplet picture still applies here. Fig. \ref{fig3} schematically shows the excitonic energy levels with and without exchange interaction included at $\pm K$ valleys. Our calculations show that all low-lying excitons are twofold degenerate at each valley, which results from the spin degeneracy and is protected by parity-time $\mathcal{PT}$ symmetry. D$_1$ and D$_4$, both coming from the transition between v$_1$ and c$_1$, are such exchange-split pairs, corresponding to triplet and singlet, respectively. D$_4$ excitons bear a strong resemblance to gray excitons in TMDCs \cite{RN8,RN9}, which can couple with out-of-plane polarized light and possess a weak oscillator strength. D$_2$ and B$_1$ arise from transition between v$_2$ and c$_1$. The level splittings due to exchange interaction are labeled as $\Delta_x^D$ (splitting between D$_1$ and D$_4$) and $\Delta_x^B$ (splitting between D$_2$ and B$_1$) in Fig. \ref{fig3}. Table II gives the exchange splittings $\Delta_x^{D/B}$ shown in Fig. \ref{fig3}, in comparison with the available experimental data. We see from Table II that, noticeably, huge exchange splitting of more than 500 meV is obtained for MnPS$_3$, which is more than ten times larger than TMDCs. Since it is sensitive to changes in the wave function overlap, the exchange energy depends on the exciton Bohr radius \cite{RN48}. Bohr radius, defined as $a_B=\varepsilon/\mu$, is governed by the static dielectric constant $\varepsilon$ and electron-hole reduced mass $\mu$. In monolayer MnPS$_3$, the significantly larger quasiparticle band gap ($E_G\approx 3.52$ eV vs. 2.8 eV in MoS$_2$ \cite{RN5}), substantially weaker screening ($\varepsilon\approx2$ vs. 8 in MoS$_2$ \cite{RE_MoS2dielec}), and notably heavier exciton reduced mass (e.g. $\mu_B(-K)\approx 4.689m_e$ vs. $0.25m_e$ of the A exciton in MoS$_2$ \cite{RE_MoS2mass}), collectively result in the observed large exciton binding energy and significantly enhanced exchange energy. 

Ultimately, the intrinsic valley polarization in MnPS$_3$, which enables direct observation of intertwined excitonic correlations and spin-valley degrees of freedom, establishes N\'{e}el-type MnPS$_3$ as a new paradigm in the field of quantum materials. Besides the disparity of valley band energy, there are also slight difference in the effective masses of the $K$ and $-K$ valleys, further dictating valley contrast in both binding energies and exchange splitting energies. As quantitatively demonstrated in Table SIV and Table II: (i) a remarkable exciton valley splitting up to $E_{D_1}(K)-E_{D_1}(-K)=123$ meV for the first dark exciton, accompanied by a $E_{B_1}(K)-E_{B_1}(-K)=-9.8$ meV for the first bright exciton, which is large enough for experiments to resolve; and (ii) pronounced valley anisotropy in excitonic exchange splitting energies, reaching significant values of $\Delta_x^D (K)-\Delta_x^D (-K)=-170$ meV and $\Delta_x^B (K)-\Delta_x^B (-K)=-45$ meV. Interestingly, the substantial valley contrast reshapes the hierarchy of excitonic states, establishing a distinct valley-dependent ordering. At $K$ valley, the lowest-energy exciton corresponds to the spin-triplet like exciton D$_2 (K)$, arising from the v$_2$ to c$_1$ transition, rather than the D$_1 (K)$ state associated with the v$_1$ to c$_1$ transition, as schematically shown in the left panel of Fig. \ref{fig3}. This seemingly counterintuitive trend is attributed to the synergistic effects of electron-hole binding strength and exchange interaction. By contrast, the $-K$ valley exhibits a distinct configuration, that is, the lowest-energy exciton D$_1 (-K)$ involves v$_1$ to c$_1$ transition with spin flip. Such valley asymmetry fundamentally modifies exciton intravalley relaxation dynamics. For $K$ valley, optical accessible B$_1$ excitons can relax to the ground state through an intraband spin-flip process. While for $-K$ valley excitons, an additional interband scattering process is required. These competing mechanisms suggest valley-dependent excitonic coherence lifetimes and thermalization behaviors, necessitating systematic investigation of the energy comparison between bright-dark splitting ($\sim$ hundreds of meV) and the characteristic phonon energies ($\hbar \omega_{ph}$ $\sim$ tens of meV \cite{RN49}) in MnPS$_3$. Establishing this quantitative correlation will elucidate how thermal population redistribution among excitonic states modulates the optoelectronic response in the future.

To demonstrate the importance of SOC effects on the valley polarization, we further perform GW-BSE excluding SOC to explore exciton effects shown in Fig. S5 in Supplemental Material \cite{SM}. In this case, the spreading of wavefunctions for the first dark and bright excitons becomes symmetrically delocalized across the $\pm K$ valleys, destroying the valley polarization. The real-space distributions remain largely unaffected, as evidenced by the minimal differences between Fig. S3(b) and Fig. S5(c) \cite{SM}. These results unambiguously demonstrate that SOC is indispensable for the emergence of spontaneous valley polarization, while real-space exciton confinement primarily stems from Coulomb interactions rather than relativistic effects.

\subsection{\label{sec:level2}C. Controlling excitonic correlation by magnetic order}
\begin{figure*}
\centering
\includegraphics[width=16cm]{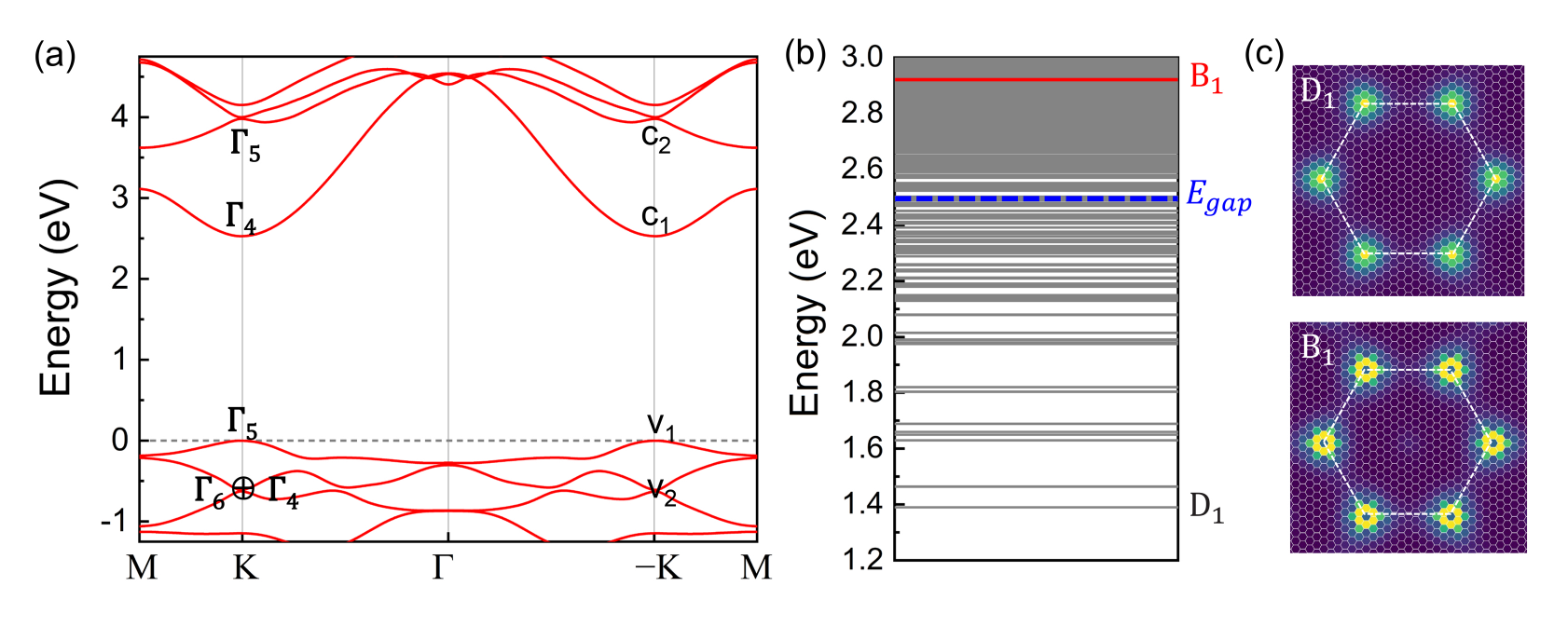}
\caption{(a) GW band structure for FM monolayer MnPS$_3$. (b) Exciton energy levels of FM MnPS$_3$ under in-plane polarized light excitation. The quasiparticle band gap is 2.52 eV. (c) Exciton spreading in $k$-space of D$_1$ exciton with $E_S=1.39$ eV (top) and B$_1$ exciton with $E_S=2.90$ eV (bottom).}
\label{fig4}
\end{figure*}

Magnetic order offers a new route to tune the exchange and correlation in 2D magnetic semiconductors, and thus their excitonic properties. When the magnetic order is switched from the N\'{e}el antiferromagnetic ground state to ferromagnetic state in MnPS$_3$, local inversion symmetry is restored, and time-reversal symmetry is broken, which lead to spin-split band structures with reduced band gap of 2.52 eV. FM phase exhibits strict spin-forbidden transition between VBM and CBM (see spin-polarized band structure in Fig. S7(a)\cite{SM}), resulting in abundant dark excitons below the quasiparticle gap as shown in Fig. \ref{fig4}(b). Furthermore, restored inversion symmetry enforces degenerate energies and identical orbital symmetries between $\pm K$ valleys. Accordingly, the excitonic ground state admits symmetric spatial distributions across K and $-$K in any basis as shown in Fig. \ref{fig4}(c), destroying the valley polarization in the FM phase. The first bright exciton B$_1$ is dominated by transitions between the majority-spin valence band and majority-spin conduction band (v$_1$ to c$_2$) near but excluding the $K$ point (Fig. \ref{fig4}(c) bottom panel). The absence of optical transitions at the $K$ point can be understood through symmetry analysis. At $K$ point, the wave-vector group can be described by $32'$ magnetic point group and its unitary subgroup corresponds to $C_3$ point group, whose character table is displayed in Table SVI in the Supplemental Material \cite{SM}. Excitons associated with v$_1$ to c$_2$ transition transform according to $\Gamma_5\otimes\Gamma_5=\Gamma_1$, which indicates its optical activity under $z$-polarized light, considering that the basis function of $\Gamma_1$ irreducible representation transforms as $z$. Away from $K$ point, the symmetry is lowered, and the optical transition is thus allowed for in-plane polarized light. Furthermore, the electronic wave function of B$_1$ primarily resides on S and P atoms (Fig. S6(b) bottom panel \cite{SM}), which matches well with the orbital characteristics of c$_2$ band dominated by the hybridization of S $p_y$ states with P $p_z$ states as evidenced by the orbital-resolved band structures in Fig. S7 \cite{SM}. The differences in band crossing, and the curvature of the bands near the band edge between the AFM and FM magnetic configurations are clearly visible. Hence, the exciton responses strongly depend on the magnetic configuration. For instance, the D$_1$ exciton has a 0.57 eV smaller binding energy for FM than AFM monolayer, which also echoes its more expanded spatial delocalization in FM by comparing Fig. S3(b) and Fig. S6(b) \cite{SM}. These findings demonstrate magnetic order as a critical knob for tailoring exciton physics in 2D magnets.

\section{\label{sec:level1}IV. SUMMARY}
In summary, we systematically explore exciton fine structures and optical properties of N\'{e}el-type antiferromagnetic monolayer MnPS$_3$ with spontaneous valley polarization. The in-gap band-spin-valley coupled exciton states show rich physics of electron-hole excitation in this 2D antiferromagnet. Notably, we have demonstrated unconventional first bright exciton involving a negative-mass hole from the second valence band, with enhanced exciton binding energies. Contrary to band-edge excitons involving S atoms $p_x\pm ip_y$ orbital, the first bright exciton exhibits significant contributions from S atoms $p_z$ orbitals, thereby offering a promising pathway for manipulation through band hybridization with adjacent layers or substrates.  Meanwhile, we reveal giant exchange energies over 10 times larger than those in TMDCs and 2D perovskites. Such large splitting energy coupled with spin-valley locking significantly modulates excitonic ordering, showing distinct dynamics properties at $\pm K$ valleys in MnPS$_3$. In addition, excitonic transitions can be drastically changed when the ground-state N\'{e}el order is fully polarized to ferromagnetic order, enabling efficient control of Coulomb correlation. Understanding and controlling the exciton fine structure in MnPS$_3$ is key for full utilization of their unique optical properties, and thus our findings of the band-spin-valley coupled exciton physics in 2D intralayer antiferromagnet should shed light on design principles for future optoelectronic, valleytronic and magneto-optical devices.

\section{\label{sec:level1}ACKNOWLEDGMENTS}
\begin{acknowledgments}
This work was supported by the National Key Research and Development Program of China (No. 2022YFA1205300), the Innovation Program for Quantum Science and Technology (Grant No. 2023ZD0300500), the Basic Science Center Project of the NSFC (Grant No. 52388201), the National Key Basic Research and Development Program of China (Grant No. 2024YFA1409100), the Shenzhen Basic Research Project (JCYJ20241202123916023) and Shenzhen Science and Technology Program (ZDSYS20230626091100001). 
\end{acknowledgments}

\section{\label{sec:level1}DATA AVAILABILITY}
The data that support the findings of this article are not publicly available. The data are available from the authors upon reasonable request.

%

\end{document}